\documentclass[10pt,twocolumn,letterpaper]{article}

\usepackage{cvpr}
\usepackage{times}
\usepackage{epsfig}
\usepackage{graphicx}
\usepackage{amsmath}
\usepackage{amssymb}

\usepackage{caption}
\usepackage{subfigure}
\usepackage{color}
\usepackage{placeins}
\usepackage{float}
\usepackage{tabularx,colortbl}
\usepackage{times}
\usepackage{epstopdf}
\usepackage{mathrsfs}
\usepackage{diagbox}
\usepackage{multirow}
\usepackage{algorithm}
\usepackage{algorithmic}
\usepackage{galois}
\usepackage{url}
\usepackage{amssymb}

\cvprfinalcopy % *** Uncomment this line for the final submission

\def\cvprPaperID{5027} % *** Enter the CVPR Paper ID here

\ifcvprfinal\fi

\begin{document}

%%%%%%%%% TITLE
\title{Learning Image and Video Compression through Spatial-Temporal Energy Compaction}

\author{Zhengxue Cheng, Heming Sun, Masaru Takeuchi, Jiro Katto\\
Department of Computer Science and Communication Engineering, Waseda University, Tokyo, Japan\\
%Tokyo 169-8555, Japan.\\
{\tt\small \{zxcheng@asagi., hemingsun@aoni., masaru-t@aoni., katto@\}waseda.jp}
}

%\and

\maketitle
\thispagestyle{empty}

%%%%%%%%% ABSTRACT
\begin{abstract}
Compression has been an important research topic for many decades, to produce a significant impact on data transmission and storage. Recent advances have shown a great potential of learning image and video compression. Inspired from related works, in this paper, we present an image compression architecture using a convolutional autoencoder, and then generalize image compression to video compression, by adding an interpolation loop into both encoder and decoder sides. Our basic idea is to realize spatial-temporal energy compaction in learning image and video compression. Thereby, we propose to add a spatial energy compaction-based penalty into loss function, to achieve higher image compression performance. Furthermore, based on temporal energy distribution, we propose to select the number of frames in one interpolation loop, adapting to the motion characteristics of video contents. Experimental results demonstrate that our proposed image compression outperforms the latest image compression standard with MS-SSIM quality metric, and provides higher performance compared with state-of-the-art learning compression methods at high bit rates, which benefits from our spatial energy compaction approach. Meanwhile, our proposed video compression approach with temporal energy compaction can significantly outperform MPEG-4 and is competitive with commonly used H.264. Both our image and video compression can produce more visually pleasant results than traditional standards.
\end{abstract}

%%%%%%%%% BODY TEXT

\section{Introduction}

%\begin{figure}[tb]
%	\centerline{\psfig{figure=visualization2.PNG,width=82.0mm} }
%	\caption{Example of one reconstruction image (\emph{kodim21}) with an approximate compression ratio of 200:1 from Kodak dataset.}
%	\label{fig:visualization2}
%\end{figure}

\begin{figure}[tb]
\centering
\subfigure[Reconstructed images \emph{kodim21} from kodak dataset.]{%$PSNR = 41.90$,
\label{fig:visualization2}
\includegraphics[width=82.0mm]{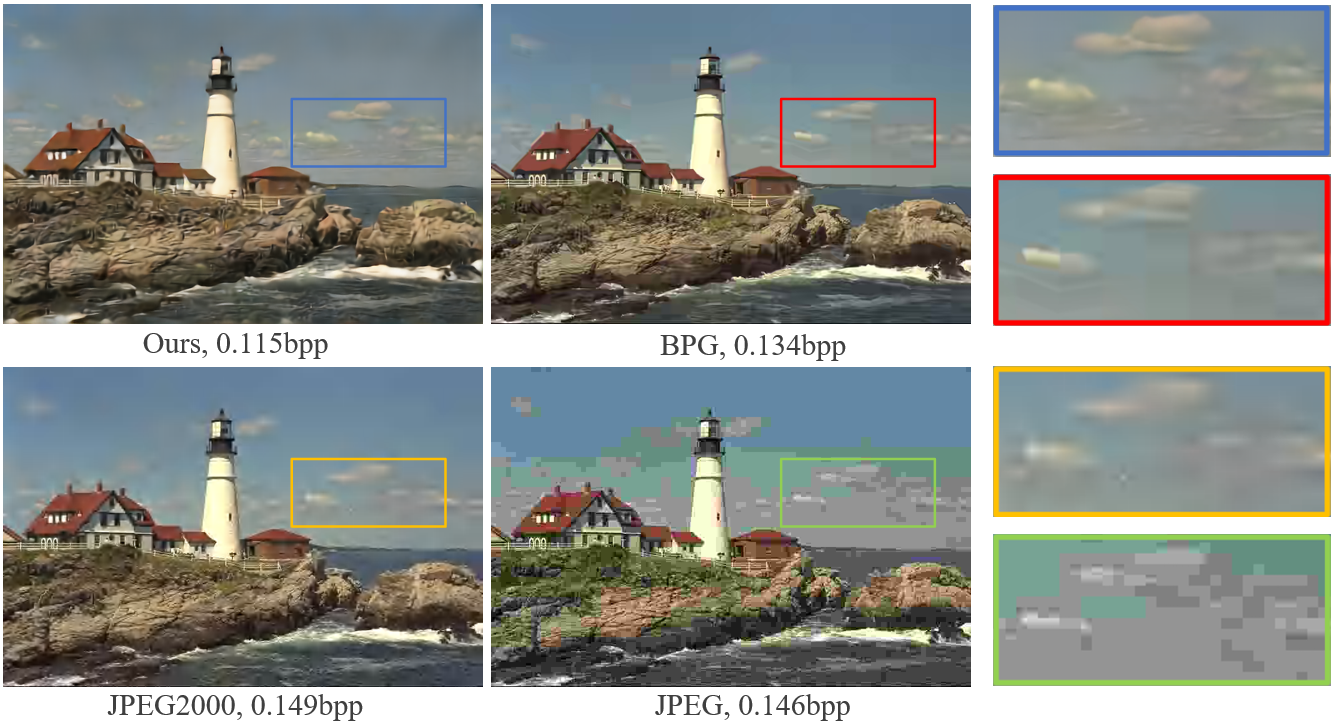}}
\subfigure[Reconstructed frame \emph{akiyo\_cif} from VTL dataset.]{%$PSNR=41.66$,
\label{fig:visualizationakiyo}
\includegraphics[width=82.0 mm]{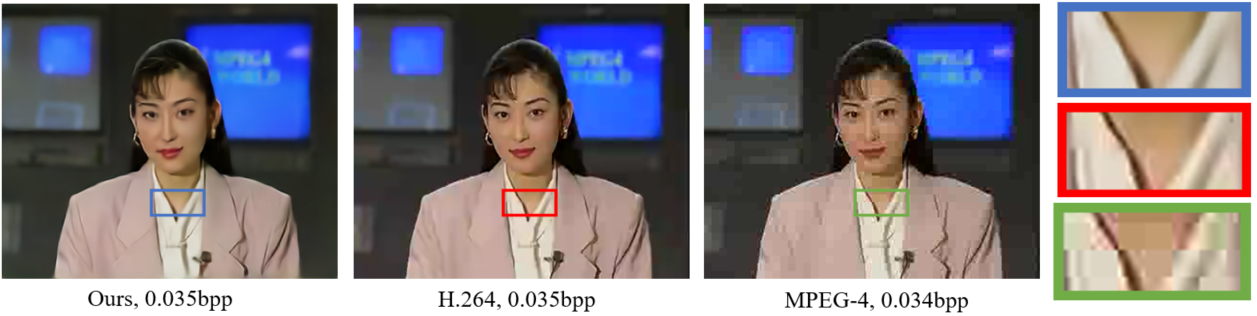}}
\caption{Visualized results of our approach and commonly used image/video compression standards. With the same bit bugets, our approach produces more visually pleasant results than conventional standards.}
\label{fig:vis}
\end{figure}

Data compression has been a significant research topic in the field of signal processing for several decades. In terms of image compression codecs, existing image compression standards, such as JPEG~\cite{IEEEexample:JPEG}, JPEG2000~\cite{IEEEexample:JPEG2000}, and BPG, which uses the intra-coded HEVC~\cite{IEEEexample:HEVC}, rely on the hand-crafted encoder-decoder pipeline. These image formats are widely used in various image applications. Conventional video coding algorithms, including H.261, MPEG-4 Part 2, commonly used standard H.264/AVC~\cite{IEEEexample:H264}, most recent standard HEVC/H.265~\cite{IEEEexample:HEVC}, have also achieved impressive performance through efforts spanning several decades. However, along with the proliferation of high-resolution images and videos, as well as the development of novel image/video formats, existing standards are not expected to be the optimal compression solution for all types of contents.

%Image compression has been a significant research topic in the field of image processing for several decades. Conventional image compression standards, such as JPEG~\cite{IEEEexample:JPEG}, JPEG2000~\cite{IEEEexample:JPEG2000} and BPG using intra-coded HEVC~\cite{IEEEexample:HEVC}, rely on the hand-crafted encoder-decoder (codec) architecture. They use the fixed transform matrices, such as discrete cosine transform (DCT) or discrete wavelet transform (DWT), together with uniform quantization and entropy coder to compress images. Good compression performance has been achieved through the efforts of several decades. However, along with the proliferation of high-resolution images and the development of more emerging image formats, existing standards are not an optimal compression solution for all types of image contents.

Recently, deep learning has been successfully applied to compression tasks. There are several potential advantages for learning compression to enhance the performance of image and video compression. First, the encoder-decoder pipeline in conventional compression standards resembles an autoencoder to learn high-level representation. Although autoencoders are basically applied to dimensionality reduction tasks~\cite{IEEEexample:autoencoder}, they are able to achieve better compression performance. Most recent learning based compression approaches, including recurrent neural networks~\cite{IEEEexample:Toderici01, IEEEexample:Toderici, IEEEexample:Nick}, convolutional neural networks~\cite{IEEEexample:Theis, IEEEexample:Balle, IEEEexample:Balle2, IEEEexample:softQuan, IEEEexample:conditional, IEEEexample:PCS, IEEEexample:HKPU} and generative adversarial networks~\cite{IEEEexample:waveone, IEEEexample:MITgan, IEEEexample:Extreme}, have all adopted the autoencoder architecture. Next, the temporal redundancy of video compression can be intuitively reduced by using learning based video prediction, generation and interpolation approaches. The works~\cite{IEEEexample:wuECCV, IEEEexample:chenVCIP} have already achieved promising results by using prediction or interpolation neural networks. The last advantage of learning based compression is that, although the development and standardization of a conventional codecs has historically spanned several years, a deep learning based compression approach can be adapted much quicker because all of the parameters in an autoencoder architecture can be learned in an automatic and unsupervised manner. Therefore, learning compression is expected to become more generalized and more efficient.

However, there are still issues to be addressed. Generally speaking, image compression exploits spatial distribution, and video compression exploits temporal distribution to learn high-level features. Good spatial-temporal energy compaction is important for high coding efficiency, according to traditional digital coding theory~\cite{IEEEexample:book}. Previous works use rate-distortion optimization, but few works analyze whether spatial-temporal energy is well-compacted or not.

In this paper, we present a convolutional autoencoder architecture with differential quantization and entropy estimation for image compression. Thereby, we propose to add an energy compaction-based penalty into loss function, to achieve higher image compression performance. Furthermore, we generalize our image compression to video compression, by adding an interpolation loop to image encoder and propose to adaptively select number of frames in one interpolation loop by analyzing temporal energy distribution.

We compare our image compression to state-of-the-art image standards and recent learning approaches. Our approach achieves significantly better MS-SSIM in comparison with the latest image compression standard BPG and outperforms state-of-the-art learning compression methods at high bit rate. On the other hand, our video compression is competitive with H.264 with MS-SSIM and produce more visually pleasant reconstructed videos.

\section{Related Work}

\noindent\textbf{Hand-crafted Compression} $\,$ Existing image compression standards, such as JPEG~\cite{IEEEexample:JPEG}, JPEG2000~\cite{IEEEexample:JPEG2000}, and BPG, which uses the intra-coded HEVC~\cite{IEEEexample:HEVC}, reply on hand-crafted module design individually. Specifically, these modules include intra prediction, discrete cosine transform or wavelet transform, quantization, and entropy coder such as Huffman coder or content adaptive binary arithmetic coder (CABAC). They design each module with multiple modes and conduct the rate-distortion optimization to determine the best mode. During the development of next-generation compression algorithms, some hybrid methods have been proposed to improve the performance, by taking advantage of both conventional compression algorithms and latest machine learning approaches, such as~\cite{IEEEexample:CLIC,IEEEexample:Yiming}.

State-of-the-art video compression algorithms, such as HEVC/H.265~\cite{IEEEexample:HEVC}, H.264~\cite{IEEEexample:H264}, MPEG-4 Part 2, incorporate the inter prediction into the encoder architecture. Inter prediction utilize the temporal similarity of neighboring frames to reduce the transmitted information. As for the order of reference frames, both H.264 and HEVC/H.265 support two configurations, that is \emph{low delay P} and \emph{random access}. \emph{low delay P} only use the previous frames as uni-directional references, while \emph{random access} allows bidirectional referencing in a hierarchical way. \emph{Random access} can achieve higher coding efficiency than \emph{low delay P}. The key technique in inter prediction is integer and fractional motion estimation using block matching and motion compensation.

\noindent\textbf{Learning Compression} $\,$ Recently, end-to-end image compression has attracted great attention. Some approaches proposed to use recurrent neural networks (RNNs) to encode the residual information between the raw image and the reconstructed images in several iterations, such as the work~\cite{IEEEexample:Toderici01, IEEEexample:Toderici} optimized by mean-squared error (MSE) or the work~\cite{IEEEexample:Nick} optimized by MS-SSIM~\cite{IEEEexample:msssim}. Some generative adversarial networks (GANs) based techniques are proposed in~\cite{IEEEexample:waveone, IEEEexample:MITgan, IEEEexample:Extreme} for high subjective reconstruction quality at extremely low bit rates. Other notable approaches include differentiable approximations of round-based quantization~\cite{IEEEexample:Theis, IEEEexample:Balle, IEEEexample:softQuan} for end-to-end training, content-aware importance map~\cite{IEEEexample:HKPU}, hyperprior entropy model~\cite{IEEEexample:Balle2} and conditional probability models~\cite{IEEEexample:conditional} for entropy estimation.

However, learning video compression has not yet been largely exploited. Only a few related works~\cite{IEEEexample:wuECCV}\cite{IEEEexample:chenVCIP} have been proposed. Wu~\emph{et al.}~\cite{IEEEexample:wuECCV} firstly proposed to use image interpolation network to predict frames except for key frames. Chen~\emph{et al.}~\cite{IEEEexample:chenVCIP} relied on traditional block based architecture to use CNN nets for predictive and residual signals. It is highly desired to further exploit learning video compression algorithms.

\begin{figure*}[tb]
\centering
\subfigure[Learning Image Compression]{%$PSNR = 41.90$,
\label{fig.overall.1}
\includegraphics[width=78.0mm]{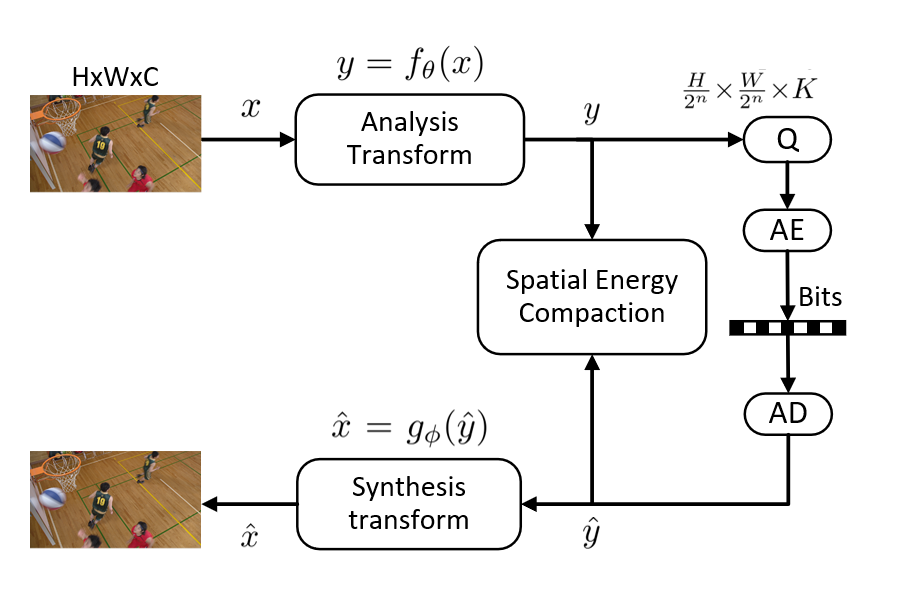}}
\subfigure[Learning Video Compression]{%$PSNR=41.66$,
\label{fig.overall.2}
\includegraphics[width=86.0 mm]{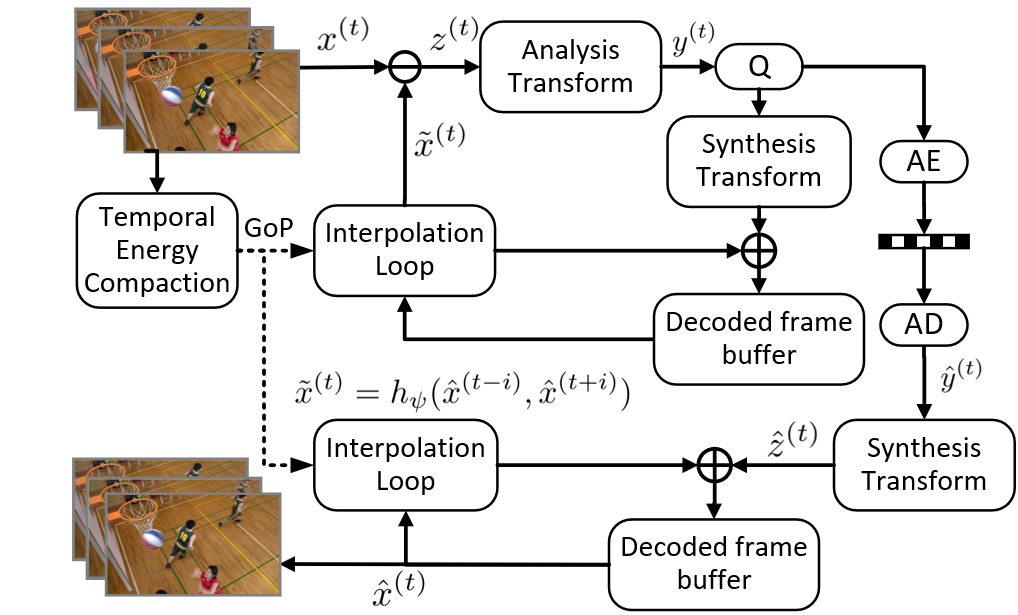}}
\caption{Overview of our proposed learning image and video compression with spatial-temporal energy compaction.}
\label{fig:overall}
\end{figure*}

\section{Proposed Method}

Our proposed learning image and video compression is illustrated in Fig.~\ref{fig:overall}, and image compression models serve for frames in video compression architecture.

\subsection{Learning Image Compression}

Given an image, a compression system can be considered as an analysis transform $f$ at the encoder side, and a synthesis transform $g$ at the decoder side, as shown in Fig.~\ref{fig.overall.1},
\begin{equation}
\begin{aligned}
y&=f_{\theta}(x) \\
\hat{x}&=g_{\phi}(\hat{y})
\end{aligned}
\end{equation}
where $x$, $\hat{x}$, $y$, and $\hat{y}$ are the original images, reconstructed images, compressed data (also called latent presentation) before quantization, and quantized compressed data, respectively. $\theta$ and $\phi$ are optimized parameters in the analysis and synthesis transforms, respectively.

\begin{figure}[b]
	\centerline{\psfig{figure=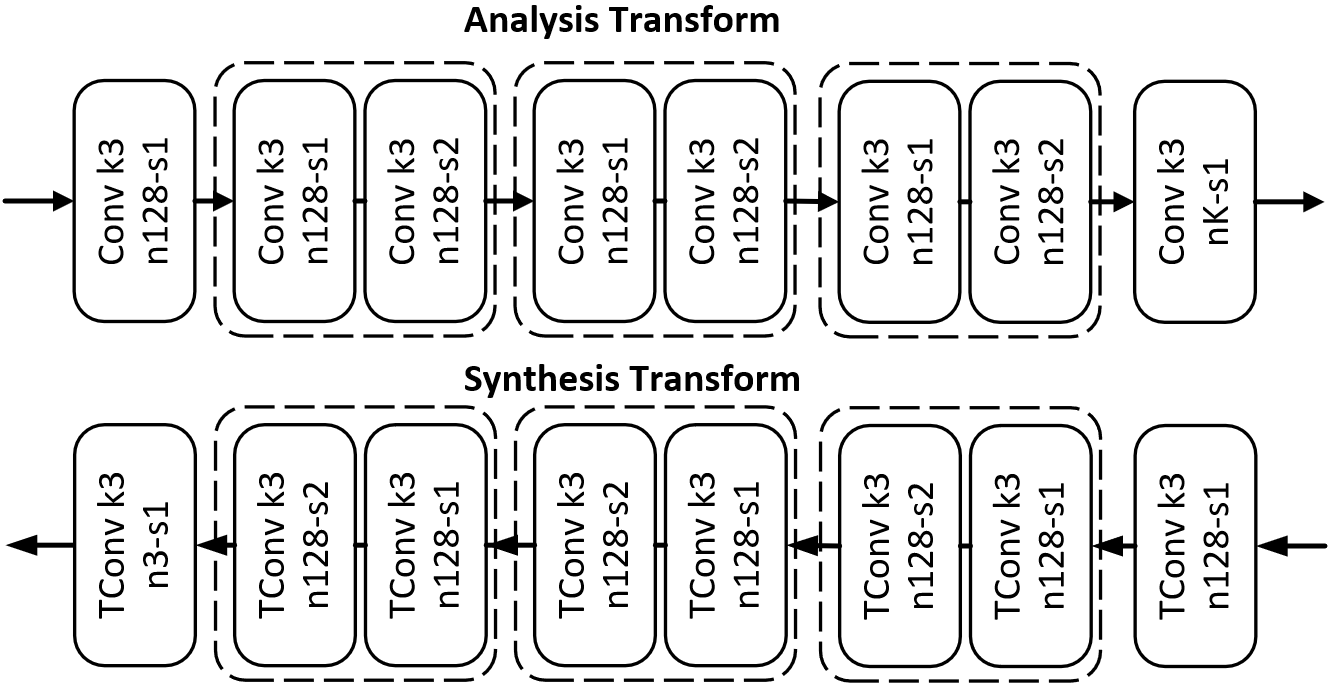,width=76.8mm} }
	\caption{Network architecture of analysis and synthesis transform, where ``k3 n128-s2'' represents a convolution layer with kernel size 3, 128 channels and a stride of 2. Tconv represents transposed convolutional layers. }
	\label{fig:unit}
\end{figure}

To obtain high-level features, the analysis and synthesis transforms can be composed into a sequence of consecutive down(up)sampling operations, which can be implemented by convolutional or transposed convolutional filter with a stride of 2. Our network architecture mainly refer to the autoencoder in~\cite{IEEEexample:Balle}, but according to ~\cite{IEEEexample:Shi}, it is pointed that super resolution is achieved more efficiently by first convolving the images and then upsampling, instead of first upsampling and then convolving. Thus we use 2 convolutional filters as one down(up)sampling unit, and the network architecture is given in Fig.~\ref{fig:unit}. Assume we have $n$ downsampling units and the number of convolutional filters in the last layer is $K$, the compressed data $y$ will have the dimension of $\frac{H}{2^{n}}\times\frac{W}{2^{n}}\times{K}$. In practice, $n=3$, $K=48$, $H$ and $W$ are set as $128$ due to memory limitation.

Based on the rate-distortion cost function in traditional codecs, the loss function is defined as follows:
\begin{equation}
\begin{aligned}
J(\theta, \phi; x) &= \lambda D(x, \hat{x})+ R(\hat{y})
\end{aligned}
\end{equation}
where $\lambda$ controls the tradeoff between the rate and distortion. $D$ represents the distortion between the original images $x$ and reconstructed images $\hat{x}$; $R$ represents the bits required to encode the quantized compressed data $\hat{y}$.

\subsubsection{Quantization and Entropy Estimation}

\begin{figure}[tb]
	\centerline{\psfig{figure=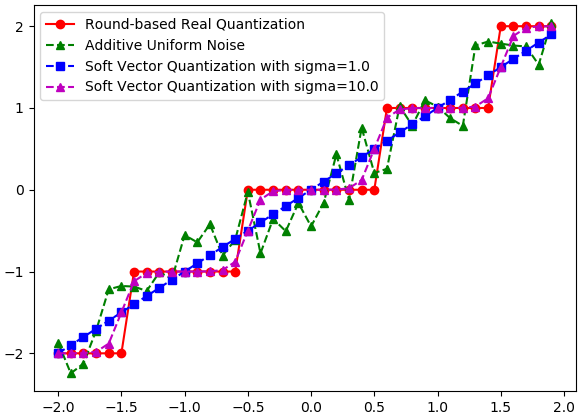,width=50.0mm} }
	\caption{Performance with different quantization methods.}
	\label{fig:quan}
\end{figure}

%Quantization is a necessary component in image compression.
In Fig.~\ref{fig.overall.1}, Q represents quantization, AE and AD represent the arithmetic encoder and arithmetic decoder, respectively. In traditional codecs, quantization is implemented by using a round function (denoted as $Q[\cdot]$), and its derivative is almost zero except at the integers. Therefore, it cannot be directly incorporated into the gradient-based optimization process. Several quantization approximations have been proposed, such as uniform noise approximation~\cite{IEEEexample:Balle} and soft vector quantization~\cite{IEEEexample:softQuan}. The approximate quantized $\hat{y}$ are shown in Fig.~\ref{fig:quan}, where soft vector quantization has a shaping parameter $\sigma$ and high $\sigma$ leads to accurate results, low $\sigma$ is good for smooth gradient propagation. In other studies~\cite{IEEEexample:Theis}\cite{IEEEexample:conditional}, the derivation was replaced in the back propagation only, but it was guaranteed that the quantized value was correct in the forward propagation. By conducting experiments, we found different quantization methods have very little effect on the compression performance. For simplicity, we used additive uniform noise approximation.

According to the Shannon theory~\cite{IEEEexample:shannon}, the rate is lower-bounded by the entropy of the discrete probability distribution of the quantized codes, as follows:
\begin{equation}
R = \mathop{\mathbb{E}}_{\tilde{y}\sim q}[-\log_{2}(p_{\tilde{y}}(\tilde{y}))]
\end{equation}
where $q$ is the actual distribution of the compressed code $\tilde{y}$ and $p_{\tilde{y}}(\tilde{y})$ is the entropy model. Thus, several entropy estimation methods have been introduced by related studies, including soft histogram based entropy estimation~\cite{IEEEexample:softQuan}, non-parametric factorized entropy model~\cite{IEEEexample:Balle2}, 3D-CNN based conditional probability model~\cite{IEEEexample:conditional}, and hyperprior based entropy model~\cite{IEEEexample:Balle2}. We used a fully factorized entropy model~\cite{IEEEexample:Balle2}, which yields promising image compression performance. Factorized entropy model produces an estimated entropy and serves for AE and AD. During the test, we can use JPEG2000 entropy coder to generate bitstream.

\subsubsection{Spatial Energy Compaction Constraint}

According to digital coding theories~\cite{IEEEexample:book}, good energy compaction property is critical for high coding efficiency performance. The analysis transform converts input $x$ into compressed data $y$ with $K$ spatial channels, which resembles a subband coding system. In a subband coding system, for any arbitrary transforms (which is not required to be non-orthogonal), the energy of $K$ spatial channels satisfies~\cite{IEEEexample:VCIP91},
\begin{equation}
\label{eq.linear}
\begin{aligned}
\sigma_{y}^{2} &= A_{k}\sigma_{x}^{2}\\
\sigma_{r}^{2} &= \sum_{k=0}^{K-1}B_{k}\sigma_{q}^{2}
\end{aligned}
\end{equation}
where $q$ is the quantization error for each spatial channel, i.e. $q = \hat{y}-y$, and $r$ is reconstructed error of images, i.e. $r = \hat{x} - x$. $\sigma^{2}$ denotes the variance of a certain data to represent the energy. $A_k$ describes the energy distribution of channels in the analysis transform, determined by $x$ and optimized parameter $\theta$; $B_k$ measures the extent of the quantization error's impact on the reconstructed error for a specified channel in the synthesis transform, determined by quantization errors and parameter $\phi$. $A_{k}$ can be easily calculated by obtaining the variance of $x$ and $y$. $B_{k}$ is determined by both the quantization errors and the parameter $\phi$ during the synthesis transform. $B_k$ can be estimated by constructing several fake compressed data $c_k$, whose $k$-th channel is all-$1$ and other channels are all-$0$. By feeding these $c_k$ as $\hat{y}$ into a given pre-trained synthesis transform individually, we stack $\sigma_{\hat{x}}^{2}$ to form $B_k$. Both $A_{k}$ and $B_{k}$ have the dimension of $K\times1$.

Based on Eq.(\ref{eq.linear}), optimum bit allocation is formulated~\cite{IEEEexample:VCIP91}. With a constant rate constraint, minimized reconstruction error is given by
\begin{equation}
\label{eq.minsigmar}
\min\{\sigma_{r}^{2}\} \propto  \prod_{i}^{K-1}(A_{k}B_{k})
\end{equation}

Detailed proof can be referred to~\cite{IEEEexample:VCIP91}. If $\prod_{i}^{K-1}(A_{k}B_{k})$ can be minimized, spatially energy can be optimally compact. Thus, we propose to add a penalty in loss function to regularize $A_k$ and $B_k$. First, we need to center the energy in a few channels as much as possible. We normalize $A_k$ by dividing its sum, therefore, normalized $A_k$ measures the energy distribution for $y_k$. For example, if $A_k[e]=0.8$ for the $e$-th channel, 80\% of energy will be distributed in the $e$-th channel. Then, we construct a penalty term by using the entropy of the energy distribution as follows:
\begin{equation}
P = \mathop{\mathbb{E}}[-A_{k}\log_{2}A_k]
\end{equation}

We add this penalty to the loss function. After several iterations, most of the energy is centered only in one or a few channels, while the other channels have little energy. We denote the channel with the largest energy as $e$. Next, we minimize the corresponding $B_k[e]$ to make $A_{k}B_{k}$ minimized, and the penalty term is defined as:
\begin{equation}
P = B_{k}[e]
\end{equation}
Finally, the loss function is defined as:
\begin{equation}
\label{eq.weighted}
J(\theta, \phi; x) = \lambda D(x, \hat{x})+ R(\hat{y}) + \beta P(A_k, B_k)
\end{equation}
where $\beta$ controls the influence of the penalty term on the loss function.

\subsection{Learning Video Compression}

Considering that a video consists of consecutive frames, a video compression system can be simply extended from image compression system as
\begin{equation}
\begin{aligned}
y^{(t)} &=f_{\theta} (x^{(t)}) \\
\hat{x}^{(t)} &= g_{\phi}(\hat{y}^{(t)})
\end{aligned}
\end{equation}
where $x^{(t)}$ represents $\{x^{(0)}, x^{(1)}, \cdot\cdot\cdot, x^{(T)}\}$, and similar notations are for $y^{(t)}, \hat{y}^{(t)}, \hat{x}^{(t)}, t\in [0,T)$. We define $T$ as group of pictures (GOPs) as HEVC/H.265~\cite{IEEEexample:HEVC}, which can be encoded and decoded independently. Two consecutive groups share the same boundary frames.

Due to the temporal similarity of neighboring frames, encoding residual information between two frames can gain more coding efficiency than encoding them separately. Thus, a more general form of video compression system is rewritten as
\begin{equation}
\begin{aligned}
\label{eq.vc}
z^{(t)} &= x^{(t)} - \tilde{x}^{(t)}_{E} \\
y^{(t)} &= f_{\theta}(z^{(t)})\\
\hat{z}^{(t)} &=g_{\phi}(\hat{y}^{(t)}) \\
\hat{x}^{(t)} &=  \hat{z}^{(t)} + \tilde{x}^{(t)}_{D}
\end{aligned}
\end{equation}
where $\tilde{x}^{(t)}_{E}$ and $\tilde{x}^{(t)}_{D}$ are predicted frame using neighboring frames at encoder and decoder side, repectively. For the first frame, there is no previous information, i.e. $\tilde{x}^{(0)}_{E} =\tilde{x}^{(0)}_{D} =0$, video compression reduces to a image compression, therefore, we call these key frames as I-frame. The block diagram of the proposed learning image compression is illustrated in Fig.~\ref{fig.overall.2}.

\subsubsection{Interpolation Loop}

The closer generated predictive frame $\tilde{x}^{(t)}$ gets to raw frames $x^{(t)}$, the fewer information $z^{(t)}$ has. Therefore, a high-quality frame interpolation network is desired. We use the latest work~\cite{IEEEexample:intp} to formulate the interpolation as a convolution $h$ over two neighboring frames as
\begin{equation}
\label{eq.intp}
\tilde{x}^{(t)} = h_{\psi}(x^{(t-i)}, x^{(t+i)})
\end{equation}
where $i$ is the distance between reference frames and generated frames. $\psi$ are optimized parameters in interpolation network $h$. More importantly, according to Eq.(\ref{eq.vc}), predicted frames should be kept the same at both encoder and decoder side to prevent the quality gap,
\begin{equation}
\tilde{x}^{(t)}_E = \tilde{x}^{(t)}_D
\end{equation}
Therefore, the encoder and decoder should see the same information equally. Then, the input for interpolation network should be reconstructed frames, not raw frames. Eq.(\ref{eq.intp}) is rewritten as
\begin{equation}
\tilde{x}^{(t)} = h_{\psi}(\hat{x}^{(t-i)}, \hat{x}^{(t+i)})
\end{equation}
We use a local interpolation loop at the encoder side to store reconstructed frames in the buffer, as Fig.~\ref{fig.overall.2}.

\subsubsection{Temporal Energy Compaction}

To further reduce the temporal redundancy, inspired by HEVC random access~\cite{IEEEexample:HEVC} and the work~\cite{IEEEexample:wuECCV}, we use a hierarchy interpolation method, which can be illustrated as
\begin{equation}
\begin{aligned}
z^{(0)} &= x^{(0)}, z^{(T)} = x^{(T)} \\
\tilde{x}^{(\frac{T}{2})} &= h_{\psi}(\hat{x}^{(0)}, \hat{x}^{(T)})\\
\tilde{x}^{(\frac{T}{4})} &= h_{\psi}(\hat{x}^{(0)}, \hat{x}^{(\frac{T}{2})}), \, \tilde{x}^{(\frac{3*T}{4})} = h_{\psi}(\hat{x}^{(\frac{T}{2})}, \hat{x}^{(T)})
\end{aligned}
\end{equation}
The hierarchy interpolation is recursively conducted until all the frames are reconstructed.

\begin{figure}[tb]
\centering
\subfigure[Sequence \emph{Akiyo} in VTL, $H_{T}$=2.673.]{%$PSNR = 41.90$,
\label{fig:ablation.1}
\includegraphics[width=58.0 mm]{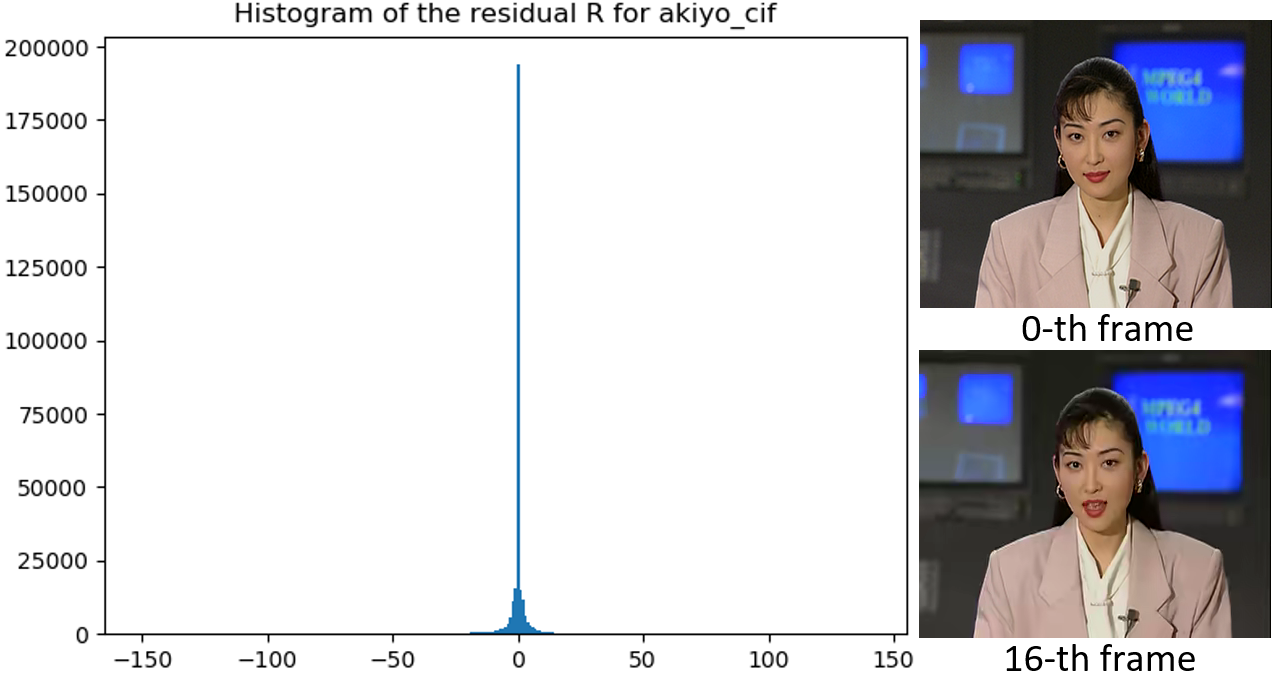}}%comparison_msssim_kodak

\subfigure[Sequence \emph{Bus} in VTL, $H_{T}$=7.999]{%$PSNR=41.66$,
\label{fig:ablation.2}
\includegraphics[width=58.0 mm]{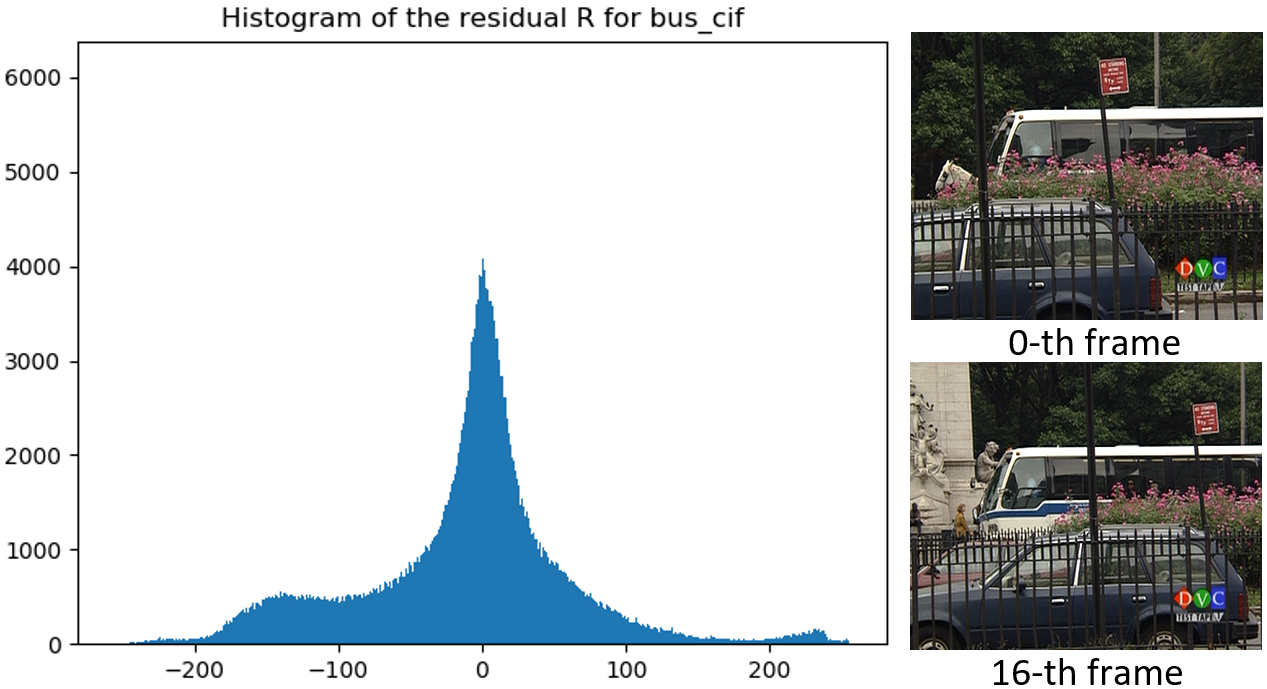}}%comparison_msssim_clic
\caption{Examples of Temporal Energy Histogram for $R_{T}$}
\label{fig:temporal}
\end{figure}

Each video contents has different motion textures, so the $T$ should be adaptively selected to fit the motion characteristic of videos, so we propose an adaptive $T$ determination strategy based on temporal energy compaction. We define the temporal motion difference in two neighboring I-frames with a proper distance $\tau$ (in our experiments $\tau=16$) as
\begin{equation}
R_{T} = x^{(\tau)} - x^{(0)}
\end{equation}
then, we consider the distribution of $R_{T}$, and calculate the entropy of $R_{t}$ as
\begin{equation}
H_{T} = \mathop{\mathbb{E}}[-P_{R_{T}}\log_{2}P_{R_{T}}]
\end{equation}
$H_{T}$ describes the motion characteristic of video sequences, as shown in Fig.~\ref{fig:temporal}. Large $H_{T}$ implies that this video has fast motion objects, while low motion videos has small $H_{T}$. Then we propose to select the $T$ using
\begin{eqnarray}T=
\label{eq.hier}
\begin{cases}
2, &U \leq H_{T} \cr
8, &L \leq H_{T} < U\cr
16, &H_{T} < L
\end{cases}
\end{eqnarray}
where $L,U$ are constants for lower and upper bounds. Low motion videos are assigned with $T=16$, that is, intermediate $(T-1)$ frames can be interpolated, without destroying the quality. In this case, $z^{(t)}$ is already small, so we send fewer I-frames to achieves temporal energy compaction. As for high motion videos, $T$ is only set as $2$, because I-frames do not provide enough information to interpolate a high-quality frame, so we remove the hierarchy interpolation to prevent the error propagation.

\section{Implementation Details}

\noindent\textbf{Dataset} To train our image compression models, we used a subset of ImageNet database~\cite{IEEEexample:ImageNet}, and cropped them into millions of $128\times 128$ samples. For testing, we used commonly used Kodak lossless image database~\cite{IEEEexample:kodak} with 24 uncompressed $768\times 512$ images. To validate the robustness of our proposed method, we also tested the proposed method on the CVPR workshop CLIC validation dataset~\cite{IEEEexample:CLICdata} with large and various resolutions up to about 2K. To test the performance of video compression, we use the widely used Video Trace Library (VTL) dataset~\cite{IEEEexample:vtl}, which includes $20$ videos with the resolution of $352\times288$ and 8 test sequences with the resolution of $832\times480$ and $416\times240$, which are commonly used by video coding standardization group with rich texture scenes and motion scenes.

\noindent\textbf{Training Details}$\quad$ To train our image compression autoencoder, the model was optimized using Adam~\cite{IEEEexample:adam} with a batch size of 16. The learning rate was maintained at a fixed value of $1\times10^{-4}$ during the training process. In the Eq.(\ref{eq.weighted}), $\beta$ was set to $0.001$. In our experiments, we add the energy compaction penalty to the model at high bit rate, and train for several $10^{5}$ iterations, and then train the model up to several $10^{6}$ iterations for each $\lambda$. By introducing different $\lambda$ to fine-tune a pre-trained autoencoder, we can obtain variable bit rates. We have found that by changing $\lambda$ with a pre-trained autoencoder, the energy distribution property will not be changed largely, as long as the initial state of the parameters in the models already has a good spatially energy compaction. Thus, we only consider the penalty for one $\lambda$ trial when training the neural network. Here, we obtained six models with $\lambda$ in the set $\{2,4,8,16,32,64\}$.

In our video compression approach, we use the pre-trained models of~\cite{IEEEexample:intp} to build our reconstruction loop. By checking the histogram of $H_{T}$ using VTL dataset, we used $L = 6.0$, $U=8.0$ in Eq.(\ref{eq.hier}) to ensure majority of sequences to select proper $T$ and averaged $T$ is equal to 8. The encoding of $z^{(t)}$ can be compressed either the trained models by our image architectures or traditional codecs. In the following results, we use BPG to encode residual for simplicity because it can provide more quality levels. 

\noindent\textbf{Measurements}$\quad$ To achieve high subjective quality, we used a popular MS-SSIM~\cite{IEEEexample:msssim} as distortion term defined by $D = 1 - \rm{MS\text{-}SSIM}(x, \hat{x})$. To measure the coding efficiency, the rate is measured in bits per pixel (bpp), and the rate-distortion (RD) curves are drawn to demonstrate their coding efficiency.

\section{Experiments}

In this section, we conduct experiments to evaluate the performance, and present comparison results.

\subsection{Ablation study}

In order to show the effectiveness of our proposed spatial-temporal energy compaction approach, we first perform the following ablation study.

\begin{figure}[tb]
\centering
\subfigure[MS-SSIM evaluated on Kodak.]{%$PSNR = 41.90$,
\label{fig:ablation.1}
\includegraphics[width=40.0 mm]{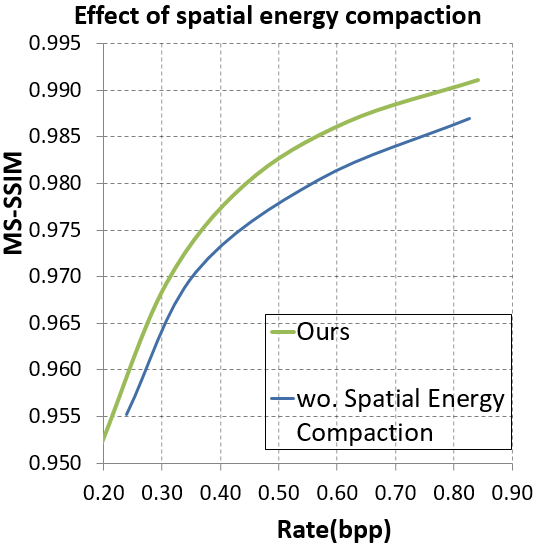}}%comparison_msssim_kodak
\subfigure[MS-SSIM evaluated on VTL.]{%$PSNR=41.66$,
\label{fig:ablation.2}
\includegraphics[width=40.0 mm]{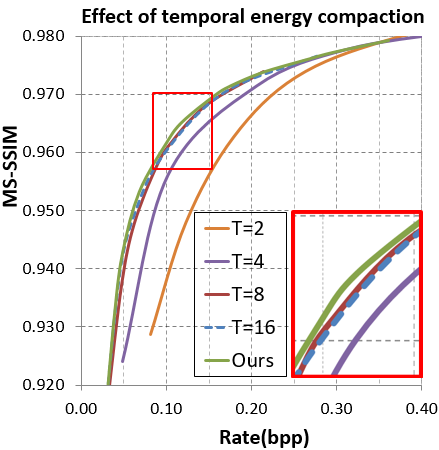}}%comparison_msssim_clic
\caption{Ablation Study.}
\label{fig:ablation}
\end{figure}

We compare the performance of our image compression with spatial energy compaction constraint to the case without energy constraint. The RD performance averaged on the Kodak dataset is presented in Fig.~\ref{fig:ablation.1}. It is observed that energy compaction constraint can help autoencoder to gain a higher coding efficiency, especially with large bit budgets.

To visualize how temporal energy compact works, we conduct the experiments of learning video compression with different $T$, as shown in Fig.~\ref{fig:ablation.2}. Along with the increasing of $T$, coding efficiency gets improved, but the performance almost saturates between $T=8$ and $T=16$. Our approach can adaptively select $T$ to achieve better rate-distortion optimization than the case with constant $T$.

\begin{figure*}[tb]
\centering
\subfigure[Kodak image dataset.]{%$PSNR = 41.90$,
\label{fig:kodakclic.1}
\includegraphics[height=42.0 mm]{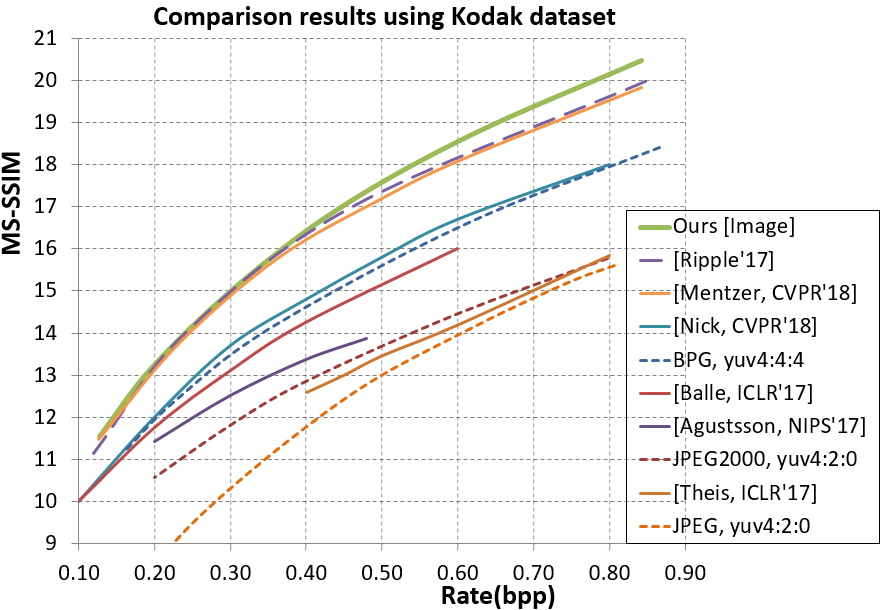}}%comparison_msssim_kodak
\subfigure[CLIC image dataset.]{%$PSNR=41.66$,
\label{fig:kodakclic.2}
\includegraphics[height=42.0 mm]{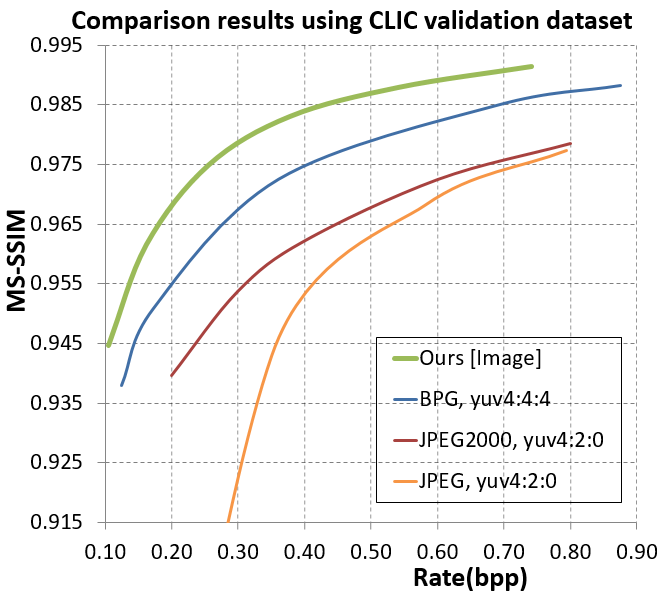}}%comparison_msssim_clic
\subfigure[VTL video dataset.]{%$PSNR=41.66$,
\label{fig:kodakclic.3}
\includegraphics[height=42.0 mm]{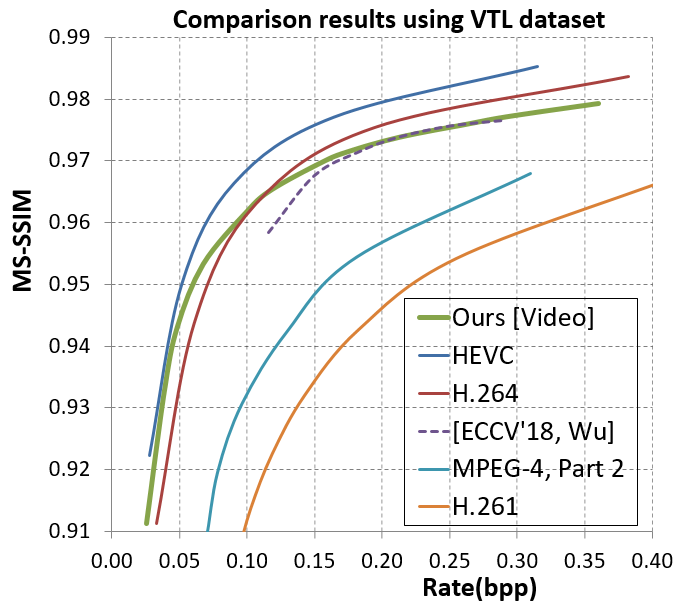}}%comparison_msssim_clic
\caption{Comparison results using different datasets.}
\label{fig:kodakclic}
\end{figure*}

\begin{figure*}[tb]
\centering
\subfigure{%$PSNR = 41.90$,
\label{fig:vc.1}
\includegraphics[height=32.0 mm]{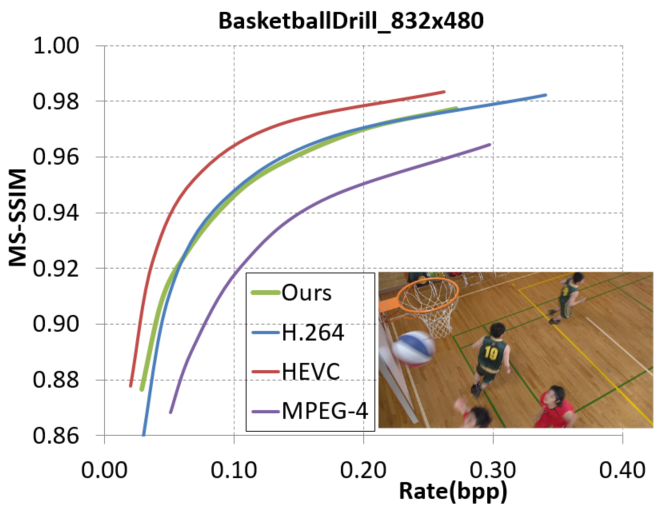}}%comparison_msssim_kodak
\subfigure{%$PSNR=41.66$,
\label{fig:vc.2}
\includegraphics[height=32.0 mm]{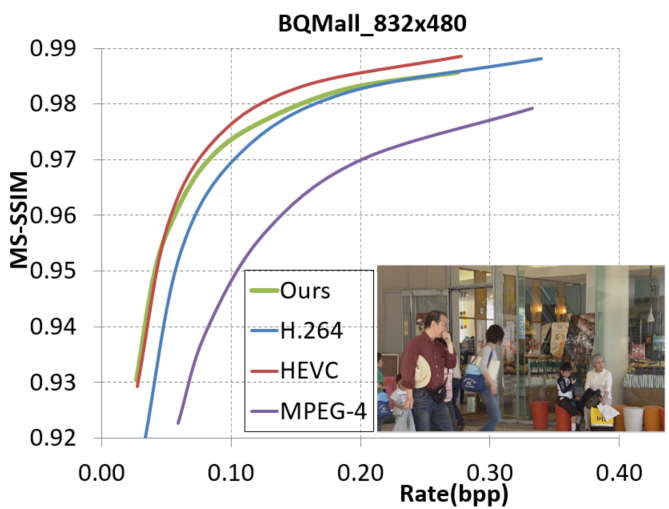}}%comparison_msssim_clic
\subfigure{%$PSNR=41.66$,
\label{fig:vc.3}
\includegraphics[height=32.0 mm]{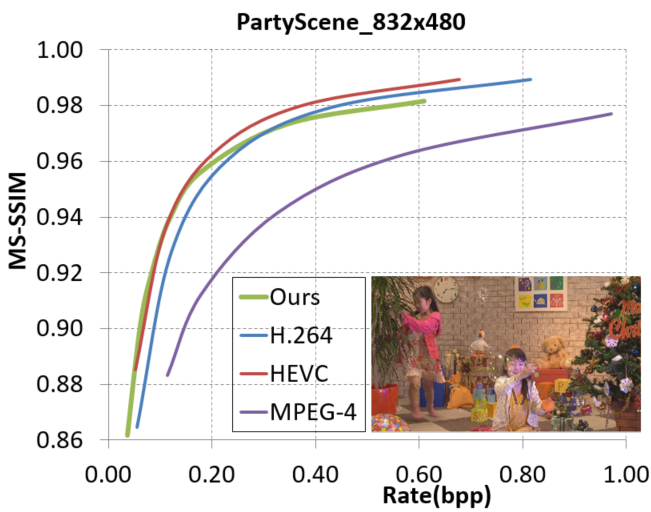}}%comparison_msssim_clic
\subfigure{%$PSNR=41.66$,
\label{fig:vc.4}
\includegraphics[height=32.0 mm]{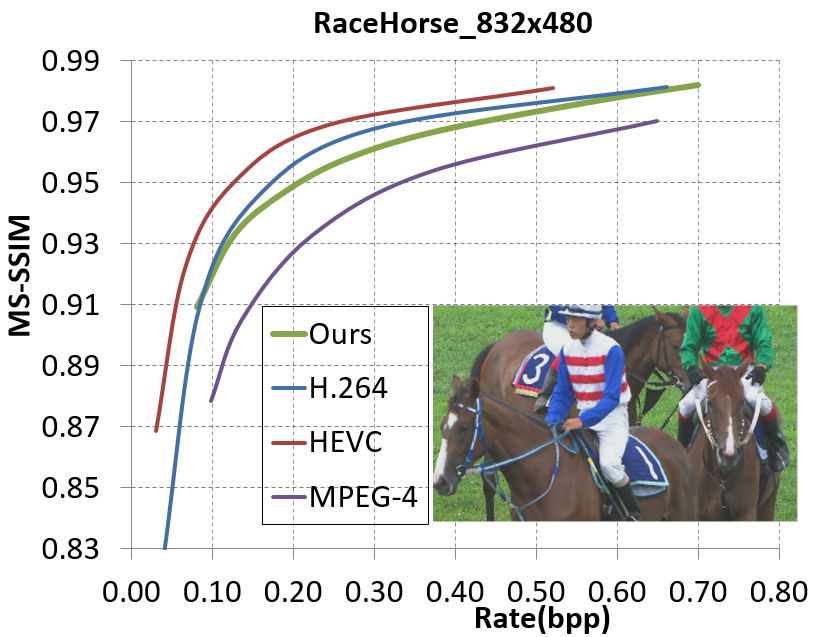}}%comparison_msssim_clic

\subfigure{%$PSNR = 41.90$,
\label{fig:vc.5}
\includegraphics[height=32.0 mm]{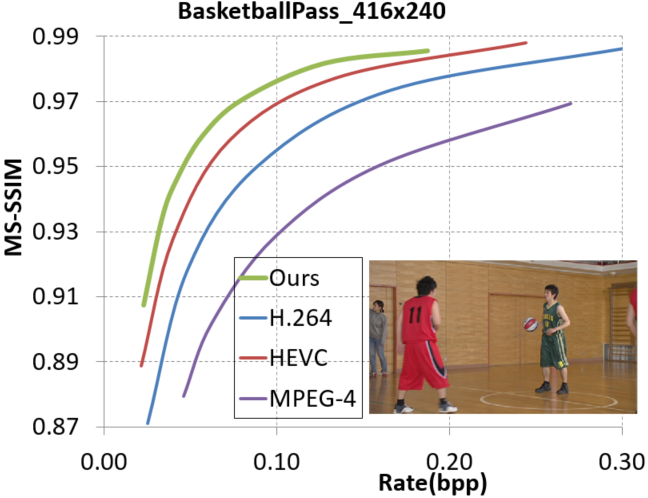}}%comparison_msssim_kodak
\subfigure{%$PSNR=41.66$,
\label{fig:vc.6}
\includegraphics[height=32.0 mm]{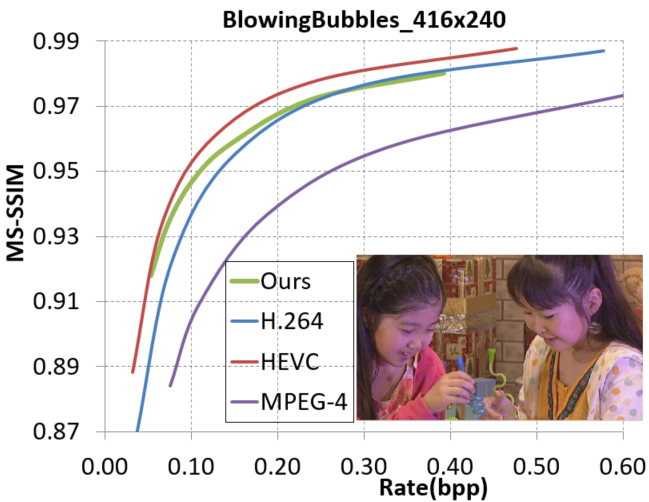}}%comparison_msssim_clic
\subfigure{%$PSNR=41.66$,
\label{fig:vc.7}
\includegraphics[height=32.0 mm]{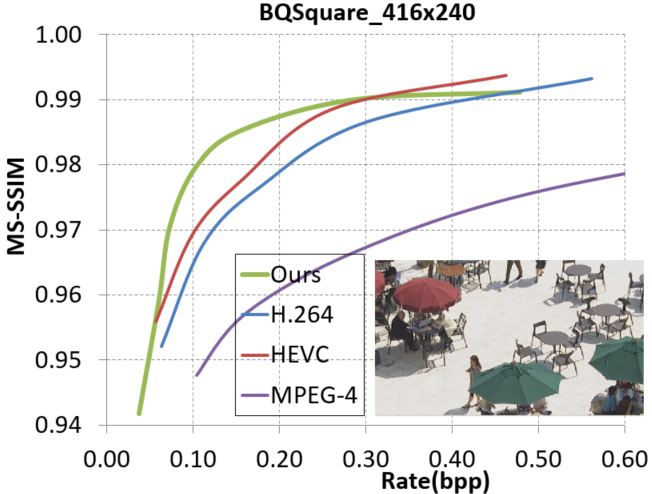}}%comparison_msssim_clic
\subfigure{%$PSNR=41.66$,
\label{fig:vc.8}
\includegraphics[height=32.0 mm]{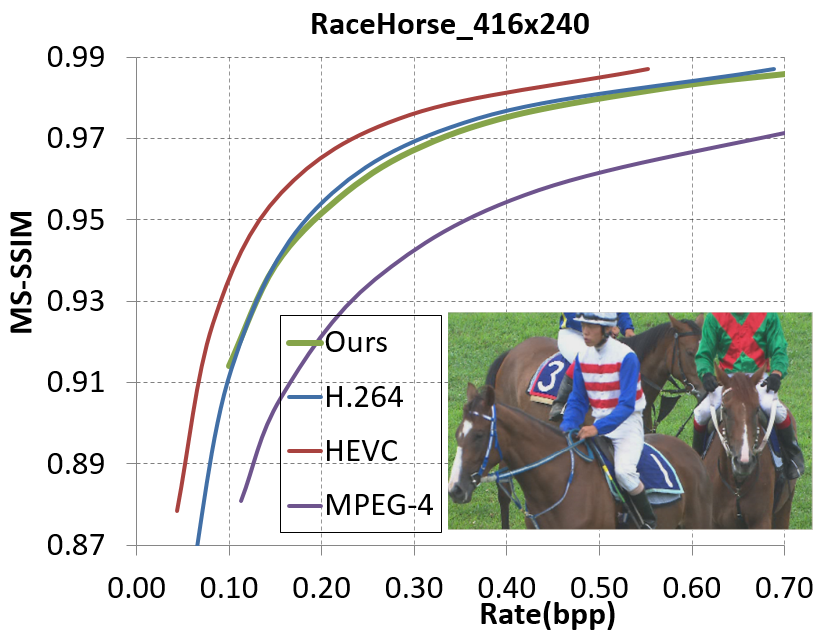}}%comparison_msssim_clic
\caption{Comparison results for each video sequence.}
\label{fig:vc}
\end{figure*}

\subsection{Performance of learning image compression}

We compare our method with well-known compression standards, and recent neural network-based learned compression methods, as shown in Fig.~\ref{fig:kodakclic.1}, where MS-SSIM is converted to decibels ($-10\log_{10}(1-$MS-SSIM$)$) to illustrate the difference clearly. For JPEG and JPEG2000, we used the official software libjpeg~\cite{IEEEexample:libjpeg} and OpenJPEG~\cite{IEEEexample:Openjpeg}, which uses the default configuration \emph{YUV420} format. The state-of-the-art image compression standard was BPG~\cite{IEEEexample:BPG}, for which we used the non-default \emph{YUV444} format refer to~\cite{IEEEexample:conditional}\cite{IEEEexample:waveone}. Because the source codes of previous neural networks based approaches were not available, we carefully traced the point in the RD curves from the respective studies of Nick \emph{et al.}~\cite{IEEEexample:Nick}, Theis \emph{et al.}~\cite{IEEEexample:Theis}, Ball\'e \emph{et al.}~\cite{IEEEexample:Balle}, and Ripple \emph{et al.}~\cite{IEEEexample:waveone}. The data in Mentzer ~\emph{et al.}~\cite{IEEEexample:conditional} were obtained from their project page. It can be observed that our method significantly outperforms Nick \emph{et al.}~\cite{IEEEexample:Nick}, Theis \emph{et al.}~\cite{IEEEexample:Theis}, Ball\'e \emph{et al.}~\cite{IEEEexample:Balle}. Our methodology is better than the work of Mentzer ~\emph{et al.}~\cite{IEEEexample:conditional} and Ripple \emph{et al.}~\cite{IEEEexample:waveone} at high bit rate, because our spatially energy compact constraint can allocate bits more efficiently, with higher bit budgets. Our proposal achieves comparable performance with Mentzers work~\emph{et al.}~\cite{IEEEexample:conditional} and Ripple \emph{et al.}~\cite{IEEEexample:waveone} at low bit rate. Currently we only use a factorized entropy model and our method does not depend on the design of entropy models. Thus, in future work, our bit allocation method can be combined with a more complicated context adaptive entropy model, such as~\cite{IEEEexample:Balle2}, to yield a better result. Fig.~\ref{fig:kodakclic.2} shows the comparison between JPEG, JPEG2000, and BPG averaged on the CLIC validation dataset. Our method outperformed JPEG, JPEG2000 and BPG significantly in terms of MS-SSIM, with high resolution images.

\begin{figure*}[tb]
	\centerline{\psfig{figure=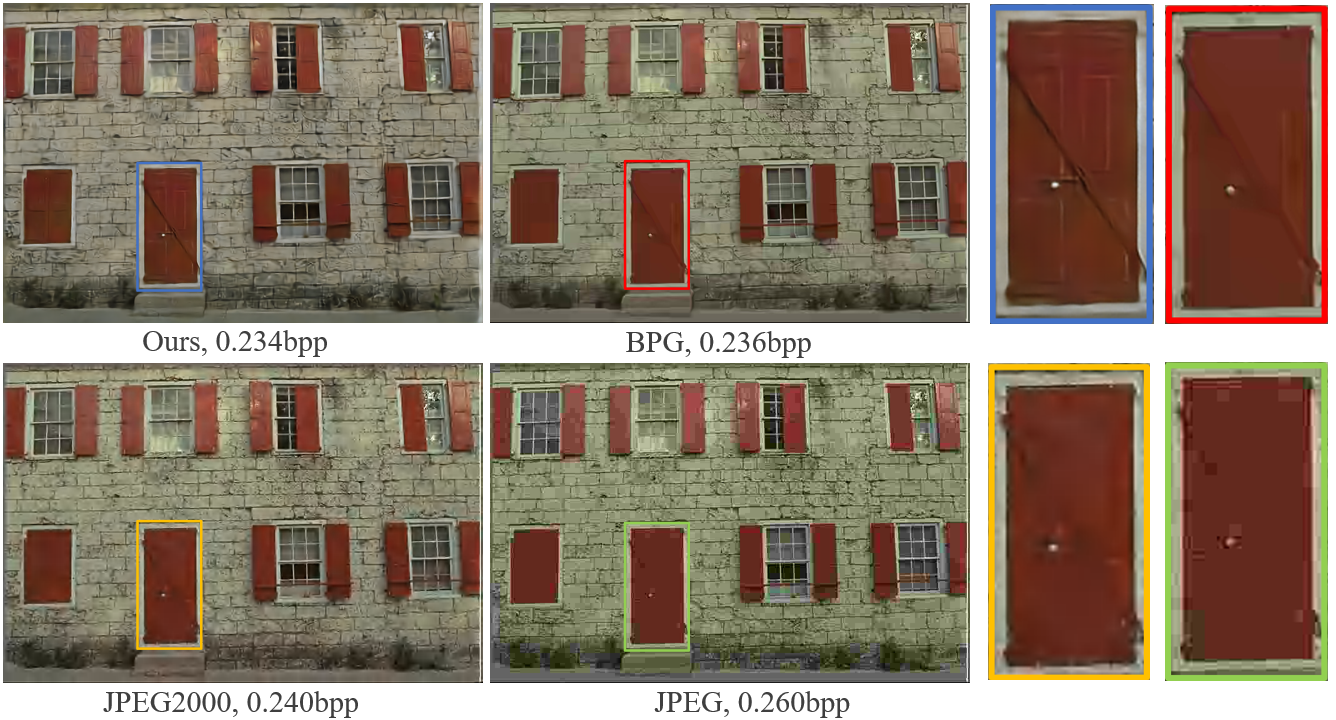,width=175.0 mm} }
	\caption{Example of one reconstruction image \emph{kodim01} from Kodak dataset.}
	\label{fig:visualization1}
\end{figure*}
\begin{figure*}[tb]
	\centerline{\psfig{figure=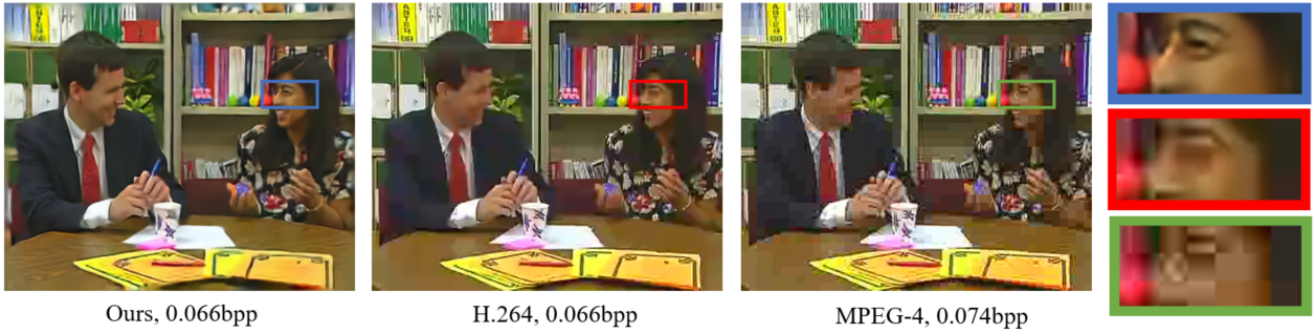,width=175.0 mm} }
	\caption{Example of one reconstruction frame in Video \emph{paris\_cif} from VTL dataset.}
	\label{fig:visualizationparis}
\end{figure*}

\subsection{Performance of learning video compression}

We compare our learning video compression with state-of-the-art video compression algorithm and recent learning video compression methods~\cite{IEEEexample:wuECCV}. For fair comparison, we use the the averaged results on each video sequences as~\cite{IEEEexample:wuECCV}. The performance using VTL is shown in Fig.~\ref{fig:kodakclic.3}. For HEVC/H.265 and H.264, we use the official software HM 16.0~\cite{IEEEexample:HM} and JM 19.0~\cite{IEEEexample:JM} with random access configuration. The GOP is set as $8$ and intra period is also $8$ to make the comparison fair. For MPEG-4 Part2 and H.261, we use FFMPEG software. The data point of \cite{IEEEexample:wuECCV} are from their original paper. It is observed that our method outperforms MPEG-4 and H.261 significantly and is competitive with H.264. Moreover, we offer a wide range of bit rates and achieve better performance even at low bit rate than the work~\cite{IEEEexample:wuECCV}, which benefits from our proposed interpolation loop and temporal energy compaction.

To cover a variety of video contents, we also test our codec using common test sequences, following the work~\cite{IEEEexample:chenVCIP}, whose results are a little worse than H.264. The RD curves of eight sequences are shown in Fig.~\ref{fig:vc}. It can be observed that our method outperforms H.264 for most sequences, and even outperforms HEVC/H.265 for sequences \emph{BasketballPass} and \emph{BQSquare}.

\subsection{Qualitative results}

We visualize some reconstructed images and videos to demonstrate qualitative performance.

The reconstructed images are shown in Fig.~\ref{fig:visualization2} and Fig.~\ref{fig:visualization1}. Fig.~\ref{fig:visualization2} shows examples \emph{kodim21} with approximately $0.12$ bpp and a compression ratio of 200:1. It can be observed that the cloud above the sea appear more natural in our reconstructed images using $0.115$ bpp less bits than BPG, JPEG2000, and JPEG. Particularly, for the BPG-encoded images, blocking artifacts occur in the sky when a large compression ratio is applied. Fig.~\ref{fig:visualization1} shows examples \emph{kodim01} under approximately $0.24$ bpp with a compression ratio of 100:1, because the raw images are a lossless PNG format with 24 bpp (8 bit for each color component). Thus, it is observed that the latch on the door is maintained well in our reconstructed images. However, the images are blurry for the BPG, JPEG2000 and JPEG reconstructed images.

Some reconstructed frames from VTL dataset are shown in Fig.~\ref{fig:visualizationakiyo} and Fig.~\ref{fig:visualizationparis}. Using the interpolation loop, the rate can be greatly saved. Clear block artifacts are observed for MPEG-4 compressed frames. Many details and shapes, such as woman's eyes in Fig.~\ref{fig:visualizationparis}, are destroyed in H.264 compressed frames. Unlike them, our approach do not have any block artifacts to produce visually pleasant results.

\section{Conclusion}

We propose learning image and video compression approach through spatial-temporal energy compaction property. Specifically, we propose to add a spatial energy compaction-based penalty into loss function in image compression models, to achieve higher performance. We generalize image compression to video compression with an interpolation loop and adaptive interpolation period selection based on entropy of temporal information.

Experimental results demonstrate that our proposed image compression outperforms BPG with MS-SSIM quality metric, and provides higher performance compared with state-of-the-art learning compression methods. Our video compression approach outperforms MPEG-4, and is competitive with commonly used H.264. Both our image and video compression can produce more visually pleasant results than traditional standards.

%\section*{Acknowledgment}


%------------------------------------------------------------------------

%{\small
%\bibliographystyle{ieee}
%\bibliography{egbib}

{\small
\begin{thebibliography}{1}

\bibitem{IEEEexample:JPEG}
G. K Wallace, \emph{``The JPEG still picture compression standard''}, IEEE Trans. on Consumer Electronics, vol. 38, no. 1, pp. 43-59, Feb. 1991.

\bibitem{IEEEexample:JPEG2000}
Majid Rabbani, Rajan Joshi, \emph{``An overview of the JPEG2000 still image compression standard''}, ELSEVIER Signal Processing: Image Communication, vol. 17, no, 1, pp. 3-48, Jan. 2002.

\bibitem{IEEEexample:HEVC}
G. J. Sullivan, J. Ohm, W. Han and T. Wiegand, \emph{``Overview of the High Efficiency Video Coding (HEVC) Standard''}, IEEE Transactions on Circuits and Systems for Video Technology, vol. 22, no. 12, pp. 1649-1668, Dec. 2012.

\bibitem{IEEEexample:H264}
T. Wiegand, G. J. Sullivan, G. Bjontegaard, A. Luthra, \emph{``Overview of the H.264/AVC Video Coding Standard''}, IEEE Transactions on Circuits and Systems for Video Technology, vol. 13, no. 7, pp. 560-576, July. 2003.

\bibitem{IEEEexample:autoencoder}
P. Vincent, H. Larochelle, Y. Bengio and P.-A. Manzagol, \emph{``Extracting and composing robust features with denoising autoencoders''}, Intl. conf. on Machine Learning (ICML), pp. 1096-1103, July 5-9. 2008.

%\bibitem{IEEEexample:TMMLi}
%S. Li et al., \emph{``Closed-Form Optimization on Saliency-Guided Image Compression for HEVC-MSP,''}, IEEE Transactions of Multimedia, Vol. 20, No. 1, Jan. 2018.

%\bibitem{IEEEexample:TMMZhu}
%S. Zhu, M. Li, C. Chen, S. Liu, and B. Zeng, \emph{``Cross-Space Distortion Directed Color Image Compression,''}, IEEE Transactions of Multimedia, Vol. 20, No. 3, March 2018.

\bibitem{IEEEexample:CLIC}
Z. Cheng, H. Sun, M. Takeuchi, J. Katto, \emph{``Performance Comparison of Convolutional AutoEncoders, Generative Adversarial Networks and Super-Resolution for Image Compression''}, CVPR Workshop and Challenge on Learned Image Compression (CLIC), pp. 1-4, June 17-22, 2018.

\bibitem{IEEEexample:Yiming}
Z. Chen, Y. Li, F. Liu, Z. Liu, X. Pan, W. Sun, Y. Wang, Y. Zhou, H. Zhu, S. Liu, \emph{``CNN-Optimized Image Compression with Uncertainty based Resource Allocation''}, CVPR Workshop and Challenge on Learned Image Compression (CLIC), pp. 1-4, June 17-22, 2018.

%\bibitem{IEEEexample:TaoDCC}
%W. Tao et al., "An End-to-End Compression Framework Based on Convolutional Neural Networks," 2017 Data Compression Conference (DCC), pp. 463-463, Snowbird, UT, 2017.


\bibitem{IEEEexample:Toderici01}
G. Toderici, S. M.O'Malley, S. J. Hwang, et al., \emph{``Variable rate image compression with recurrent neural networks''}, arXiv: 1511.06085, 2015.

\bibitem{IEEEexample:Toderici}
G, Toderici, D. Vincent, N. Johnson, et al., \emph{``Full Resolution Image Compression with Recurrent Neural Networks''}, IEEE Conf. on Computer Vision and Pattern Recognition (CVPR), pp. 1-9, July 21-26, 2017.

\bibitem{IEEEexample:Nick}
Nick Johnson, Damien Vincent, David Minnen, et al., \emph{``Improved Lossy Image Compression with Priming and Spatially Adaptive Bit Rates for Recurrent Networks''}, arXiv:1703.10114, pp. 1-9, March 2017.

\bibitem{IEEEexample:Theis}
Lucas Theis, Wenzhe Shi, Andrew Cunninghan and Ferenc Huszar, \emph{``Lossy Image Compression with Compressive Autoencoders''}, Intl. Conf. on Learning Representations (ICLR), pp. 1-19, April 24-26, 2017.

\bibitem{IEEEexample:Balle}
J. Balle, Valero Laparra, Eero P. Simoncelli, \emph{``End-to-End Optimized Image Compression''}, Intl. Conf. on Learning Representations (ICLR), pp. 1-27, April 24-26, 2017.

\bibitem{IEEEexample:Balle2}
Johannes Balle, D. Minnen, S. Singh, S. J. Hwang, N. Johnston, \emph{``Variational Image Compression with a Hyperprior''}, Intl. Conf. on Learning Representations (ICLR), pp. 1-23, 2018. \url{https://tensorflow.github.io/compression/}

\bibitem{IEEEexample:softQuan}
E. Agustsson, F. Mentzer, M. Tschannen, L. Cavigelli, R. Timofte, L. Benini, L. V. Gool, \emph{``Soft-to-Hard Vector Quantization for End-to-End Learning Compressible Representations''}, Neural Information Processing Systems (NIPS) 2017, arXiv:1704.00648v2.

\bibitem{IEEEexample:conditional}
F. Mentzer, E. Agustsson, M. Tschannen, R. Timofte, L. V. Gool, \emph{``Conditional Probability Models for Deep Image Compression''}, IEEE Conf. on Computer Vision and Pattern Recognition (CVPR), June 17-22, 2018. \url{https://github.com/fab-jul/imgcomp-cvpr}

\bibitem{IEEEexample:PCS}
Z. Cheng, H. Sun, M. Takeuchi, J. Katto, \emph{``Deep Convolutional AutoEncoder-based Lossy Image Compression''}, Picture Coding Symposium, pp. 1-5, June 24-27, 2018.

\bibitem{IEEEexample:HKPU}
M. Li, W. Zuo, S. Gu, D. Zhao, D. Zhang, \emph{``Learning Convolutional Networks for Content-weighted Image Compression''}, IEEE Conf. on Computer Vision and Pattern Recognition (CVPR), June 17-22, 2018.

\bibitem{IEEEexample:waveone}
Ripple Oren, L. Bourdev, \emph{``Real Time Adaptive Image Compression''}, Proc. of Machine Learning Research, Vol. 70, pp. 2922-2930, 2017.

\bibitem{IEEEexample:MITgan}
S. Santurkar, D. Budden, N. Shavit, \emph{``Generative Compression''}, Picture Coding Symposium, June 24-27, 2018.

\bibitem{IEEEexample:Extreme}
E. Agustsson, M. Tschannen, F. Mentzer, R. Timofte, and L. V. Gool, \emph{``Generative Adversarial Networks for Extreme Learned Image Compression''}, arXiv:1804.02958.

\bibitem{IEEEexample:wuECCV}
C-Y Wu, N. Singhal, P. Krahenbuhl, \emph{``Video Compression through Image Interpolation''}, 15th European Conference on Computer Vision, September 8 ¨C 14, 2018.

\bibitem{IEEEexample:chenVCIP}
T. Chen, H. Liu, Q. Shen, T. Yue, X. Cao, and Z. Ma. \emph{``Deepcoder: A deep neural network based video compression''}. 2017 IEEE Visual Communications and Image Processing (VCIP), pp. 1¨C4, Dec 2017.

\bibitem{IEEEexample:CLICdata}
Workshop and Challenge on Learned Image Compression, CVPR2018, \url{http://www.compression.cc/challenge/}

\bibitem{IEEEexample:Shi}
W. Shi, J. Caballero, F. Huszar, et al. \emph{``Real-time single image and video super-resolution using an efficient sub-pixel convolutional neural network''}, Intl. IEEE Conf. on Computer Vision and Pattern Recognition, June 26-July 1, 2016.

\bibitem{IEEEexample:book}
N.S. Jayant and P. Noll, \emph{``Digital coding of waveforms''}, Englewood Cliffs NJ, Prentice-Hall, 1984.

\bibitem{IEEEexample:VCIP91}
J.Katto and Y.Yasuda: \emph{``Performance Evaluation of Subband Coding and Optimization of Its Filter Coefficients''}, SPIE Visual Communication and Image Processing, Nov.1991.

%\bibitem{IEEEexample:residual}
%K. He, X. Zhang, S. Ren and J. Sun, \emph{``Deep Residual Learning for Image Recognition''}, arXiv.1512.03385, 2015.

\bibitem{IEEEexample:adam}
D. P. Kingma and J. Ba, \emph{``Adam: A method for stochastic optimization''}, arXiv:1412.6980, pp.1-15, Dec. 2014.

\bibitem{IEEEexample:ImageNet}
J. Deng, W. Dong, R. Socher, L. Li, K. Li and L. Fei-Fei, \emph{``ImageNet: A Large-Scale Hierarchical Image Database''}, IEEE Conf. on Computer Vision and Pattern Recognition, pp. 1-8, June 20-25, 2009.

\bibitem{IEEEexample:kodak}
Kodak Lossless True Color Image Suite, Download from http://r0k.us/graphics/kodak/

\bibitem{IEEEexample:msssim}
Z. Wang, E. P. Simoncelli and A. C. Bovik, \emph{``Multiscale structural similarity for image quality assessment''}, The 36-th Asilomar Conference on Signals, Systems and Computers, Vol.2, pp. 1398-1402, Nov. 2013.

%\bibitem{IEEEexample:bdrate}
%G. Bjontegaard, \emph{``Calculation of Average PSNR Differences between RDcurves''}, ITU-T VCEG, Document VCEG-M33, Apr. 2001.

\bibitem{IEEEexample:libjpeg}
JPEG official software libjpeg, https://jpeg.org/jpeg/software.html

\bibitem{IEEEexample:Openjpeg}
JPEG2000 official software OpenJPEG, \\
https://jpeg.org/jpeg2000/software.html

\bibitem{IEEEexample:BPG}
BPG Image Format, https://bellard.org/bpg/

%\bibitem{IEEEexample:ISM}
%Z. Cheng, M. Takeuchi, J. Katto, \emph{``A Pre-Saliency Map Based Blind Image Quality Assessment via Convolutional Neural Networks''}, IEEE Intl. Symposium on Multimedia, pp. 1-6, Dec. 11-13, 2017.

\bibitem{IEEEexample:shannon}
C. E. Shannon, \emph{``A Mathematical Theory of Communication''}, The Bell System Technical Journal, Vol. 27, pp. 379-423, July, 1948.

\bibitem{IEEEexample:intp}
S. Niklaus, L. Mai and F. Liu, \emph{``Video Frame Interpolation via Adaptive Separable Convolution''}, IEEE International Conference on Computer Vision (ICCV) 2017.

\bibitem{IEEEexample:vtl}
Video Trace Library. \url{http://trace.eas.asu.edu/index.html}

\bibitem{IEEEexample:HM}
K. McCann, C. Rosewarne, B. Bross, M. Naccari, K. Sharman, G. J. Sullivan, \emph{``High Efficiency Video Coding (HEVC) Test Model 16 (HM 16) Encoder Description''}, Document JCTVC-R1002, Sapporo, Jul. 2014. \url{https://hevc.hhi.fraunhofer.de/svn/svn_HEVCSoftware/}

\bibitem{IEEEexample:JM}
A. M.Tourapis, K. Suhring, G. Sullivan, \emph{``H.264/14496-10 AVC Reference Software Manual''}, Document JVT-AE010, London, UK, 28 June- 3 July 2009. \url{http://iphome.hhi.de/suehring/tml/download/}


%\bibitem{IEEEexample:normalizingflow}
%D. J. Rezende, S. Mohaned, \emph{``Variational Inference with Normalizing Flows''}, arXiv:1505.05770v6.

\end{thebibliography}
}
%}

\clearpage
\section{Supplementary Material}

\subsection{Proof of Spatial Energy Constraint}

%\section{Proof of Optimum Bit Allocation}
In Section 3.1.2, we propose a spatial energy compaction constraint. The detailed proof for this proposal is given in the following.

Let $\alpha_{k} = \frac{N_{k}}{N}$, where $N_{k}$ and $N$ are the total number of inputs and that of $y_{k}(n)$, respectively. Our autoencoder network consist of three downsampling units, so $\alpha_{k} = \frac{1}{8}$. Refer to~\cite{IEEEexample:VCIP91}, the optimum bit allocation problem is described as follows: under the constant rate constraint
\begin{equation}
\sum_{k=0}^{K-1}\alpha_{k}R_{k} = R(const)
\end{equation}
, minimize
\begin{equation}
\sigma_{r}^{2} = \sum_{k=0}^{K-1}B_{k}\sigma_{q_k}^{2}
\end{equation}
where $y$, $q_k$ has K channels, so we denote them as $y_k$ and $q_k$. $R_{k}$ is bit rate for the k-th channel. By substituting the approximating relationship~\cite{IEEEexample:book}
\begin{equation}
\label{eq.errorlinear}
\sigma_{q_k}^{2} \simeq \epsilon^{2}2^{-2R}\sigma_{y_k}^{2}
\end{equation}
where $\epsilon$ is a constant depending on the input characteristics. Using the Lagrangian multiplier method, let
\begin{equation}
L =  \sum_{k=0}^{K-1}B_{k}\epsilon^{2}2^{-2R}\sigma_{y_k}^{2}
\end{equation}
and let $\frac{dL}{d\sigma_{r}^{2}} = 0$, we can get the minimum value of the reconstruction error variance as
\begin{equation}
min\{\sigma_{r}^{2}\} =  \prod_{i}^{K-1}(\frac{A_{k}B_{k}}{\alpha_{k}})^{\alpha_{k}}\cdot \epsilon^{2}2^{-2R}\sigma_{x}^{2}
\end{equation}
where $\epsilon^{2}2^{-2R}\sigma_{x}^{2}$ and $\alpha_k$ are constants. This equation can be rewritten as
\begin{equation}
\label{eq.minsigmar}
\min\{\sigma_{r}^{2}\} \propto  \prod_{i}^{K-1}(A_{k}B_{k})
\end{equation}
The physical meaning of this equation is to describe the compression capability of neural networks. Usually the large energy channel easily have large quantization error, and the all-zero channel will have no quantization error, which is proved by Eq. (\ref{eq.errorlinear}). If $\prod_{i}^{K-1}(A_{k}B_{k})$ is minimized, it implies the quantization errors on a large energy channel (i.e., $A_k$ is large) have little impact on the final reconstruction quality, because $B_k$ should be small for this channel, which leading to high quality reconstruction. Meanwhile, in an almost-all-zero channel ($A_k$ is small), the quantization error can have large influence on the final reconstruction quality, because $B_k$ can be either large or small, whereas all-zero channel almost have no quantization error. The above analysis is our theoretical basis for our proposal.

%Besides, the coding gain of subband coding system is given by
%\begin{equation}
%G_{SBC} =  \frac{1}{\prod_{i}^{K-1}(\frac{A_{k}B_{k}}{\alpha_{k}})^{\alpha_{k}}}
%\end{equation}

\subsection{PSNR results}

\begin{figure}[htb]
	\centerline{\psfig{figure=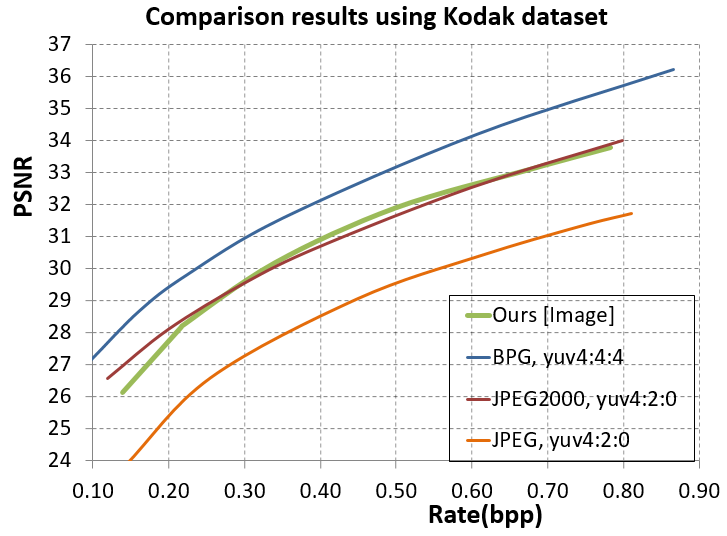,width=50.8mm} }
	\caption{Results on PSNR. }
	\label{fig:psnr}
\end{figure}

PSNR evaluation is added in Figure~\ref{fig:psnr} by optimizing MSE, where our models are comparable with JPEG2000.
%Optimizing MS-SSIM leads to better perceptual quality, proven by many studies e.g. [Ripplef17], [CVPRf18, Mentzer], [CVPRf18, Nick]. Thatfs why these papers and ours mainly discuss MS-SSIM.

\end{document}